\documentclass[11pt,fleqn]{article}

\usepackage{amsfonts}
\usepackage{amsmath}
\usepackage{amssymb}
\usepackage{mathtools}
\usepackage{mathrsfs}
\usepackage{enumerate}
\usepackage{bbm}
\usepackage{graphicx,epsfig,amsthm,color}
\usepackage{pstricks}
\usepackage{subfig}
\usepackage{graphicx}
\usepackage{units}

\usepackage{caption}
\usepackage{booktabs}
\usepackage{textcomp}
\usepackage{array}
\usepackage{cite}
\usepackage{cancel}

\date{\today}

\newcolumntype{z}[1]{>{\RaggedRight\hspace{0pt}}p{#1}}
\newcolumntype{w}[1]{>{\RaggedRight\hspace{0pt}}p{#1}}
\newcolumntype{v}[1]{>{\Centering\hspace{0pt}}p{#1}}

\def \la {\lambda}

\def \a {\alpha}

\def\be{\begin{equation}}
\def\ee{\end{equation}}
\def\bea{\begin{eqnarray}}
\def\eea{\end{eqnarray}}

\def\be{\begin{equation}}
\def\ee{\end{equation}}
\def\bea{\begin{eqnarray}}
\def\eea{\end{eqnarray}}

\def\erp2{{\rm e}^{2\rho}}
\def\erm2{{\rm e}^{-2\rho}}
\def\er4{{\rm e}^{4\rho}}

\def\be{\begin{equation}}
\def\ee{\end{equation}}
\def\bea{\begin{eqnarray}}
\def\eea{\end{eqnarray}}

\def\m0{m_{\nu_{0,i}}}

\def\T0{T_{\nu_0}}

\newcommand{\half}{\frac{1}{2}}

\newcommand{\beqa}{\begin{eqnarray}}
\newcommand{\eeqa}{\end{eqnarray}}
\newcommand{\bpr}{\begin{problem}}
\newcommand{\epr}{\end{problem}}
\newcommand{\bcent}{\begin{center}}
\newcommand{\ecent}{\end{center}}
\newcommand{\bfig}{\begin{figure}}
\newcommand{\efig}{\end{figure}}
\newcommand{\bpc}{\begin{picture}}
\newcommand{\epc}{\end{picture}}

\newcommand{\nnb}{\nonumber}
\newcommand{\reflef}{(\ref}

\renewcommand{\and}{A_{0}^{\nu ,D}(s)}

\newcommand{\bee}{\begin{equation}}

\def\beq{\begin{eqnarray}}
\def\eeq{\end{eqnarray}}

\newcommand{\Ga}{\Gamma}

\newcommand{\lmd}{\lambda}
\newcommand{\bright}{\begin{flushright}}
\newcommand{\eright}{\end{flushright}}
\newcommand{\bminip}{\begin{minipage}}
\newcommand{\eminip}{\end{minipage}}

\DeclareCaptionLabelSeparator{mysep}{\hspace{3pt}:\hspace{3pt}}
\DeclareCaptionLabelFormat{mypiccap}{Fig.\hspace{3pt}{#2}}
\DeclareCaptionLabelFormat{mytabcap}{Tab.\hspace{3pt}{#2}}

\begin{document}

\date{}
\title{
\vskip 2cm {\bf\huge Chameleonic dilaton, nonequivalent frames, and the cosmological constant problem in quantum string theory}\\[0.8cm]}

\author{
{\sc\normalsize Andrea Zanzi\footnote{Email: zanzi@th.physik.uni-bonn.de}\!\!}\\[1cm]
{\normalsize Via Pioppa 261, 44123 Pontegradella (Ferrara) -  Italy}\\}
 \maketitle \thispagestyle{empty}
\begin{abstract}
{The chameleonic behaviour of the String theory dilaton is
suggested. Some of the possible consequences of the chameleonic
string dilaton are analyzed in detail. In particular, (1) we
suggest a new stringy solution to the cosmological constant
problem and (2) we point out the non-equivalence of different
conformal frames at the quantum level. In order to obtain these
results, we start taking into account the (strong coupling) string
loop expansion in the string frame (S-frame), therefore the
so-called {\it form factors} are present in the effective action.
The correct Dark Energy scale is recovered in the Einstein frame
(E-frame) without unnatural fine-tunings and this result is robust
against all quantum corrections, granted that we {\it assume} a
proper structure of the S-frame form factors in the strong
coupling regime. At this stage, the possibility still exists that
a certain amount of fine-tuning may be required to satisfy some
phenomenological constraints. Moreover in the E-frame, in our
proposal, all the interactions are switched off on cosmological
length scales (i.e. the theory is IR-free), while higher
derivative gravitational terms might be present locally (on short
distances) and it remains to be seen whether these facts clash
with phenomenology. A detailed phenomenological analysis is
definitely necessary to clarify these
points. \\
}
\end{abstract}
This article has been
accepted for publication in Physical Review D.\\ The link to the
entry page of the journal is: http://prd.aps.org/ \\ Copyright (2010) by
the American Physical Society.

\clearpage

\newpage \thispagestyle{empty} \begin{minipage}[t]{14.4 cm}
\hspace*{-6.5em}

\mbox{}\\[9.4em]
\hspace*{-15.1em}
\mbox{}\\[4.1em]

{\hfill To the memory of my father Paolo,}

{\hfill with love and gratitude.}

 \end{minipage}
\mbox{}\\[-.0 em]
\clearpage

\tableofcontents
\newpage

\section{Introduction}

Our Universe is currently in a state of accelerated expansion
\cite{Riess:1998cb, Perlmutter:1998np, Spergel:2003cb,
deBernardis:2000gy}. Many possibilities have been discussed in the
literature to account for this acceleration (for a review see for
example \cite{Copeland:2006wr}). Among them, we can mention: 1) a
modification of General Relativity at large distances
\cite{Dvali:2000hr}; 2) a backreaction of properly smoothed-out
inhomogeneities \cite{Rasanen:2003fy, Kolb:2004am, Kolb:2005da};
3) a negative-pressure Dark Energy (DE) fluid. As far as the last
case is concerned, the nature of the DE is far from clear and its
contribution to the cosmic energy budget is comparable, at
present, with that of the dark matter component (DM), although
their dynamical evolution during the cosmological history can be
totally different ("coincidence problem"). DE could be a
cosmological constant (for a review see \cite{Nobbenhuis:2006yf})
or it could have a dynamic behaviour. Let us focus our attention
on this last possibility. On the one hand, the standard scenario
for dynamical DE is given by a scalar field which is rolling over
a (almost flat) potential (i.e. the scalar field is ultralight).
On the other hand, there are reasons to maintain a non-trivial
coupling between the scalar field and matter, for instance: a) to
solve, at least partially, the coincidence problem, a direct
interaction between DM and DE has been discussed
\cite{Amendola:1999er, Amendola:2000uh, TocchiniValentini:2001ty,
Comelli:2003cv, Pietroni:2002ey, Huey:2004qv, Amendola:2004ew,
Gasperini:2003tf}; b) string theory suggests the presence of
scalar fields (dilaton and moduli) that can be coupled to matter
with a strength comparable to (or even higher than) the
gravitational strength. Consequently, we would like to allow a
direct interaction between matter and an ultralight scalar field.
However this could be phenomenologically dangerous (violations of
the equivalence principle, time dependence of couplings - for
reviews see \cite{Uzan:2002vq, Fischbach:1999bc}).

One possible way-out is to consider "chameleon scalar fields"
\cite{Khoury:2003aq, {Khoury:2003rn}}, namely scalar fields
coupled to matter (including the baryonic one) with gravitational
(or even higher) strength and with a mass dependent on the density
of the environment. On cosmological distances, where the densities
are very tiny, the fields are ultralight and they can roll on
cosmological time scales. On the Earth instead, the density is
much higher and the field acquires a large enough mass (i.e.
stabilization). In other words, the physical properties of this
field, including its value, vary with the environment, thus the
name chameleon.

Recently, Y. Fujii suggested a connection between the cosmological
constant problem and a surprising globally massless but locally
massive dilaton (i.e. a dual mass) \cite{Fujii:2002sb} in the
framework of Scalar-Tensor theories of gravitation (for a review
see \cite{Fujii:2003pa}). In the scenario proposed by Fujii, the
dilaton $\sigma$, the (pseudo)-Nambu-Goldstone boson of broken
scale-invariance, is split into a background part $\sigma_b$
and a fluctuating part $\sigma_f$. Surprisingly, $\sigma_f$ can
acquire a large mass in the process of quantization, while
$\sigma_b$ can remain (almost) massless. Moreover, the effects of
$\sigma_f$ on $\sigma_b$ ({\it dilatonic backreaction}) play a
major role: the contribution to the vacuum energy coming from
quantum diagrams with dilatons in the external legs is (naturally)
too large and the difference with the cosmological constant scale
is, in the proposal of Fujii, the observed effect of the dilatonic
backreaction. In other words, Fujii suggested to exploit the dilatonic
backreaction as counterterm in the renormalization of the dilatonic
vacuum energy. However, several points are unsatisfactory about this approach:
\begin{itemize}
\item the theoretical origin of the model is not discussed.

 \item
A globally massless but locally massive dilaton can be accepted
and it is not forbidden by the renormalization program. It would
be rewarding to show that (1) this result {\it must} be obtained
and (2) it admits a simple description.

\item It would be rewarding to connect the dilatonic
backreaction mentioned by Fujii with the cosmological (i.e. metric) backreaction.

\item A detailed mechanism
to suppress the contribution to the vacuum energy coming from
quantum diagrams with dilatons in the external legs is missing. In other words, in the proposal of Fujii
we know that the dilatonic backreaction exists, but we do {\it not} know whether it is effective as counterterm for the dilatonic vacuum energy.
Moreover, Fujii carried on his
calculations only for the diagrams with one and two external dilatons: the
remaining diagrams (present of course in infinite number) have not
been calculated.
\item The contribution to the vacuum energy coming from matter fields,
gauge fields and gravitons is not discussed.

\item The origin of the correct DE scale is unknown;
\end{itemize}
In our analysis, we will {\it extend} the Fujii's model and we will overcome all its drawbacks:\\
1) we point out the stringy origin of our model. In particular, we
start considering the S-frame string action in the {\it strong}
coupling regime with a constant dilaton and, after a conformal
transformation to the E-frame, we will obtain a string dilaton
$\sigma$ running (on cosmological distances)
towards the region of {\it weak} coupling.\\
2) We suggest that a conceivable interpretation of the dual mass
of the dilaton is simply a chameleonic behaviour of the field.
This will be our {\it final result} in the effective potential of
the theory in the E-frame and it provides a simple description of
the dual mass.
Remarkably, this result {\it must} be obtained in our approach.\\
3) We establish a connection between the dilatonic and the metric backreactions.\\
4) We show that scale-invariance is almost restored on
cosmological distances (in the Einstein frame). In this way on
cosmological distances we protect, on the one hand, the mass of
the dilaton, on the other hand, the cosmological constant.
Our argument is valid at all orders in perturbation theory. Naturally this symmetry principle guarantees that the dilatonic backreaction {\it must} suppress the dilatonic vacuum energy (i.e. the renormalization of the dilatonic vacuum energy must be successful). Moreover, we suggest in this article to exploit the metric backreaction as a counterterm in the renormalization of the gravitational vacuum energy. Once again, our symmetry principle guarantees that the metric backreaction {\it must} suppress the gravitational vacuum energy. As far as matter and gauge fields are concerned, we will show that their contribution to the cosmological constant is properly suppressed.\\
5) The correct DE scale is recovered in the E-frame without
unnatural fine-tunings, granted that we {\it assume} a proper
structure of the string loop corrections in the S-frame in the
strong coupling regime. However, at this stage, the possibility
still exists that a certain amount of fine-tuning may be required
to satisfy some phenomenological constraints. Moreover, the model
in the E-frame is IR-free, while higher derivative gravitational
terms might be present locally (on short distances) and it remains
to be seen whether these facts clash with phenomenology. A
detailed phenomenological analysis is definitely
necessary to clarify these points.\\

It must be stressed that solving these problems in the framework
of String Theory led us to a totally new solution to the
cosmological constant problem: in our proposal the dilaton in the Einstein
frame is parametrizing the amount of (scale) symmetry of the
problem. Therefore, the chameleonic behaviour of the field implies
that Particle Physics is the standard one only {\it locally}. All
the usual contributions to the vacuum energy (from supersymmetry [SUSY] breaking,
from axions, from electroweak symmetry breaking...) are extremely
large with respect to the meV-scale only {\it locally}, while on
cosmological distances they are suppressed.

The chameleonic behaviour of the string dilaton is relevant, not
only because it is a new stringy way to deal with crucial problems
like dilaton stabilization and cosmological constant problem, but
also for experimental reasons. Indeed, in the present period a
very important source of comparison between high energy physics
theories and experimental data comes from the Large Hadron
Collider\footnote{For a review of possible extensions of the
Standard Model - SM - of particle physics see
\cite{Kazakov:2006kp}. For an overview of the testable new physics
at the LHC see \cite{Nath:2010zj}.}. However, we think that it is
very important to dedicate attention also to experiments which can
test physics beyond the SM in alternative ways with respect to the
"main road" given by accelerator physics. The present paper
reflects this approach: there is the possibility of searching for
chameleon fields in current experiments, most notably in the
optical set-up of GammeV \cite{Chou:2008gr, Upadhye:2009iv}.
Consequently, the chameleonic behaviour of the string dilaton is
relevant to establish a connection between String Theory and
experiments.

One last remark is in order. Typically different conformal frames
are equivalent at the classical level and this result is
well-established in the literature (see for example
\cite{Catena:2006bd}). In our analysis we will point out that this
equivalence is lost at the quantum level and we will select the
E-frame as the physical one. For further details on the
(non)-equivalence of different conformal frames the reader is
referred to \cite{Nojiri:2001pd, Fujii:2007qv, Alvarez:2001qj} and
references therein.

About the organization of this paper, in section \ref{stsection}
we describe Scalar-Tensor theories of gravitation; our model will
be briefly summarized in section \ref{modello}; the
scale-invariant part of our lagrangian will be further discussed
in section \ref{SI}; we will describe our solution to the
cosmological constant problem in section \ref{CC}; we will touch upon
the non-equivalence of different conformal frames
at the quantum level in section \ref{forum}. In the final
section we will draw some concluding remarks.

\setcounter{equation}{0}
\section{Scalar-Tensor theories of gravitation}
\label{stsection}

In this section we will briefly review some aspects of
Scalar-Tensor (ST) theories of gravitation following
\cite{Fujii:2003pa, Fujii:2004bg}.

\subsection{Jordan-Brans-Dicke models}

The Lagrangian of the original ST model by Jordan-Brans-Dicke (JBD) can be written in the form:
\begin{equation}
{\cal L}_{\rm JBD} = \sqrt{-g}\left(\half \xi \phi^2 R -\epsilon
\half g^{\mu\nu}\partial_\mu\phi \partial_\nu\phi +L_{\rm matter}
\right). \label{bsl1-4}
\end{equation}
$\xi$ is a dimensionless constant and $\epsilon=\pm1$ (in
particular $\epsilon=+1$ corresponds to a normal field having a
positive energy, in other words, not to a ghost). The convention
on the Minkowskian metric is (-,+,+,+). The first term on the
right-hand side is called "nonminimal coupling term" (NM), it is
unique to the ST theory and it replaces the Einstein-Hilbert term
(EH) in the standard theory:
\begin{equation}
{\cal L}_{\rm EH} = \sqrt{-g}\frac{1}{16\pi G}R. \label{bsl1-5}
\end{equation}
If we compare this last formula with the NM-term, we infer that in
this theory the gravitational constant $G$ is replaced by an
``effective gravitational constant" defined by
\begin{equation}
\frac{1}{8\pi G_{\rm eff}}= \xi \phi^2, \label{bsl1-6}
\end{equation}
which is spacetime-dependent through the scalar field $\phi(x)$.

We stress that Jordan admitted the scalar field to be included in
the matter Lagrangian $L_{\rm matter}$, whereas Brans and Dicke
(BD) assumed not. For this reason the name ``BD model'' seems
appropriate to the assumed {\it absence} of $\phi$ in $L_{\rm
matter}$.

In this paper we will be particularly interested in string theory,
supposed to be one of the most promising theoretical models of
unification, where a scalar field, often called dilaton (a
spinless partner of the tensor metric field in higher dimensional
spacetime) appears with the same coupling as had been shown by the
ST theory. Twenty years after the pioneering works by Jordan,
Fierz, Brans and Dicke we re-discover ST theory with a "top-down"
approach suggested by string theory.

\subsection{Conformal transformation}

\subsubsection{Scale transformation (Dilatation)}

Let us start with a global scale transformation in curved
spacetime, namely:

\begin{equation}
g_{\mu\nu}\rightarrow  g_{*\mu\nu}=\Omega^2
g_{\mu\nu},\quad\mbox{or}\quad g_{\mu\nu}=\Omega^{-2} g_{*\mu\nu},
\label{bsl1-45}
\end{equation}
where $\Omega$ is a constant, from which follows also
\begin{equation}
 g^{\mu\nu}=\Omega^{2} g^{*\mu\nu},\quad \mbox{and}\quad \sqrt{-g}=
 \Omega^{-4}\sqrt{-g_*}.
\label{bsl1-46}
\end{equation}

If we have only massless fields or particles, we have no way to
provide a fixed length scale, we then have a scale invariance or
dilatation symmetry. Had we considered a fundamental field or
particle having a nonzero mass $m$, the inverse $m^{-1}$ would
have provided a fixed length or time standard and the
above-mentioned invariance would have been consequently broken.

To implement this idea, let us introduce a real free {\it massive}
scalar field $\Phi$ (not to be confused with the dilaton), as a
representative of matter fields:
\begin{equation}
{\cal L}_{\rm matter}= \sqrt{-g}\left( -\half (\partial \Phi)^2
-\half m^2\Phi^2 \right), \quad (\partial \Phi)^2 \equiv
g^{\mu\nu}(\partial_\mu\Phi)(\partial_\nu\Phi). \label{bsl1-44}
\end{equation}
We then find
\begin{eqnarray}
{\cal L}_{\rm matter}&=&  \Omega^{-4}\sqrt{-g_*}\left(-\half
\Omega^2(\partial \Phi)^2 -\half m^2\Phi^2 \right) \nnb\\
&=& \sqrt{-g_*}\left(-\half (\partial_* \Phi_*)^2 -\half
\Omega^{-2}m^2\Phi_*^2 \right),\nnb\\
\quad \mbox{with}\quad \Phi_*=\Omega^{-1}\Phi. \label{bsl1-47}
\end{eqnarray}

Notice that we defined $\Phi_*$ primarily to leave the kinetic
term form invariant except for putting the $*$ symbol everywhere.
On the other hand, the mass term in the last equation breaks scale
invariance.

\subsubsection{Conformal transformation (Weyl rescaling)}

The global scale transformation in curved spacetime as discussed
above may be promoted to a {\it local} transformation by replacing
the constant parameter $\Omega$ by a local function $\Omega(x)$,
an arbitrary function of $x$.  This defines a conformal
transformation, or sometimes called Weyl rescaling:
\begin{equation}
 g_{\mu\nu}\rightarrow g_{*\mu\nu}=\Omega^{2}(x) g_{\mu\nu},\quad
 \mbox{or}\quad ds^2 \rightarrow ds^2_* = \Omega^{2}(x)ds^2.
\label{bsl1-51}
\end{equation}
According to the last equation, we are considering a local change
of units, not a coordinate transformation.  The condition for
invariance is somewhat more complicated than the global
predecessors.

Let us see how the ST theory is affected by the conformal
transformation.  We start with
\begin{equation}
\partial_{\mu}g_{\nu\lambda}=\partial_{\mu}\left( \Omega^{-2}g_{*\nu\lambda}
\right)=\Omega^{-2}\partial_{\mu}g_{*\nu\lambda} -2\Omega^{-3}\partial_{\mu}\Omega
g_{*\nu\lambda}=\Omega^{-2}\left( \partial_{\mu}g_{*\nu\lambda}
-2f_{\mu}g_{*\nu\lambda} \right), \label{bsl1-52}
\end{equation}
where $f=\ln\Omega, f_\mu =\partial_\mu f, f_*^\mu =
g_*^{\mu\nu}f_\nu$. We then compute
\begin{equation}
\Gamma^{\mu}_{\hspace{.3em}\nu\lambda}= \half g^{\mu\rho}\left(
    \partial_{\nu} g_{\rho\lambda}+\partial_{\lambda} g_{\rho\nu}
    -\partial_{\rho} g_{\nu\lambda} \right)
=\Gamma^{\mu}_{*\hspace{.1em}\nu\lambda}-\left(
f_{\nu}\delta^{\mu}_{\lambda}
  +f_{\lambda}\delta^{\mu}_{\nu}-f_{*}^{\mu}g_{*\nu\lambda}
\right), \label{bsl1-53}
\end{equation}
reaching finally
\begin{equation}
R=\Omega^2\left( R_{*}+6\Box_{*}f - 6
g_{*}^{\mu\nu}f_{\mu}f_{\nu}\right). \label{bsl1-54}
\end{equation}

Using this in the first term on the right-hand side of
\reflef{bsl1-4}) with $F(\phi) =\xi\phi^2$, we obtain
\begin{equation}
{\cal L}_1=\sqrt{-g} \half F(\phi) R =\sqrt{-g_{*}}\half
F(\phi)\Omega^{-2} \left( R_{*}+6\Box_{*}f - 6
g_{*}^{\mu\nu}f_{\mu}f_{\nu}\right). \label{bsl1-55}
\end{equation}
We may choose
\begin{equation}
F\Omega^{-2}=1, \label{bsl1-55a}
\end{equation}
so that the first term on the right-hand side goes to the standard
EH term.  We say that we have moved to the Einstein conformal
frame (E frame). We have
\begin{equation}
\Omega=F^{1/2},\quad\mbox{then } f= \ln\Omega,\hspace{1em}
f_{\mu}=\partial_\mu f=\frac{\partial_\mu\Omega}{\Omega}=\half
\frac{\partial_{\mu}F}{F} =\half\frac{F'}{F}\partial_{\mu}\phi,
\label{bsl1-56}
\end{equation}
where $F'\equiv dF/d\phi$. The second term on the right-hand side
of \reflef{bsl1-55}) then goes away by partial integration, while
the third term becomes
$-\sqrt{-g_*}(3/4)(F'/F)^2g_*^{\mu\nu}\partial_\mu\phi
\partial_\nu\phi$.  This term is added to the second term on the
right-hand side of \reflef{bsl1-4}) giving the kinetic term of
$\phi$:
\begin{equation}
-\half\sqrt{-g_{*}}\Delta
g^{\mu\nu}_{*}\partial_{\mu}\phi\partial_{\nu}\phi,
\quad\mbox{with}\quad \Delta= \frac{3}{2}\left( \frac{F'}{F}
\right)^2 +\epsilon\frac{1}{F}. \label{bsl1-57}
\end{equation}

If $\Delta >0$, we define a new field $\sigma$ by
\begin{equation}
\frac{d\sigma}{d\phi}= \sqrt{\Delta},\quad\mbox{hence}\quad
\sqrt{\Delta}\partial_\mu\phi =\frac{d\sigma}{d\phi}
\partial_\mu\phi = \partial_\mu\sigma, \label{bsl1-58}
\end{equation}
thus bringing \reflef{bsl1-57}) to a canonical form
$-(1/2)\sqrt{-g_{*}}g^{\mu\nu}_{*}\partial_{\mu}\sigma\partial_{\nu}\sigma$.
If $\Delta <0$, the opposite sign in the first expression of
\reflef{bsl1-57})  propagates to the sign of the preceding
expression, implying a ghost.

By using the explicit expression of $F(\phi)$ we find
\begin{equation}
\Delta = \left( 6+\epsilon\xi^{-1} \right)\phi^{-2} \equiv
\zeta^{-2}\phi^{-2}, \label{bsl1-59}
\end{equation}
which translates the condition $\Delta >0$ into $\zeta^2 >0$.  We
further obtain
\begin{equation}
\frac{d\sigma}{d\phi}=\zeta^{-1}\phi^{-1},\quad\mbox{hence}\quad
\zeta\sigma = \ln \left(\frac{\phi}{\phi_{0}}
\right),\quad\mbox{or}\quad
 \phi=\xi^{-1/2}e^{\zeta\sigma},
\label{bsl1-60}
\end{equation}
reaching also
\begin{equation}
\Omega =e^{\zeta\sigma}= \sqrt{\xi} \phi. \label{bsl1-60a}
\end{equation}
We finally obtain the lagrangian in the E frame:
\begin{equation}
{\cal L}_{\rm JBD}=\sqrt{-g_{*}}\left( \half R_{*} - \half
    g^{\mu\nu}_{*}\partial_{\mu}\sigma\partial_{\nu}\sigma
    +L_{\rm *{\rm matter}}  \right).
\label{bsl1-61}
\end{equation}

In the next section we will describe our model.

\section{The model}
\label{modello}

Our starting point is the string-frame, low-energy, gravi-dilaton
effective action, to lowest order in the $\a'$ expansion, but
including dilaton-dependent loop (and non-perturbative)
corrections, encoded in a few  ``form factors" $\psi(\phi)$,
$Z(\phi)$, $\alpha{(\phi)}$, $\dots$, and in an effective dilaton
potential $V(\phi)$ (obtained from non-perturbative effects). In
formulas (see for example \cite{Gasperini:2001pc} and references therein):
\bea S &=& -{M_s^{2}\over 2} \int d^{4}x \sqrt{-  g}~
\left[e^{-\psi(\phi)} R+ Z(\phi) \left(\nabla \phi\right)^2 +
{2\over M_s^{2}} V(\phi)\right]
\nonumber \\
&-& {1\over 16 \pi} \int d^{4}x {\sqrt{-  g}~  \over
\alpha{(\phi)}} F_{{\mu\nu}}^{2} + \Ga_{m} (\phi,  g, \rm{matter})
\label{3} \eea Here $M_s^{-1} = \la_s$ is the fundamental
string-length parameter and $F_{\mu\nu}$ is the gauge field
strength of some fundamental grand unified theory (GUT) group ($\a(\phi)$ is the
corresponding gauge coupling). We imagine having already
compactified the extra dimensions and having frozen the corresponding
moduli at the string scale.

Since the form factors are {\it unknown} in the strong coupling
regime, we are free to {\it assume} that the structure of these
functions in the strong coupling region implies an S-frame
Lagrangian composed of two different parts: 1) a scale-invariant
Lagrangian ${\cal L}_{SI}$. This part of our lagrangian has
already been discussed in the literature by Fujii in references
\cite{Fujii:2002sb, Fujii:2003pa}; 2) a Lagrangian which
explicitly violates scale-invariance ${\cal L}_{SB}$.

In formulas we write:

\beq {\cal L}={\cal L}_{SI} + {\cal L}_{SB}, \label{Ltotale}\eeq where the
scale-invariant Lagrangian is given by:

\begin{equation}
{\cal L}_{\rm SI}=\sqrt{-g}\left( \half \xi\phi^2 R -
    \half\epsilon g^{\mu\nu}\partial_{\mu}\phi\partial_{\nu}\phi -\half g^{\mu\nu}\partial_\mu\Phi \partial_\nu\Phi
    - \frac{1}{4} f \phi^2\Phi^2 - \frac{\lambda_{\Phi}}{4!} \Phi^4
    \right).
\label{bsl1-96}
\end{equation}
$\Phi$ is a scalar field representative of matter fields,
$\epsilon=-1$, $\zeta \simeq 1$, $f<0$ and $\lambda_{\Phi}>0$.
One may write also terms like $\phi^3 \Phi$, $\phi \Phi^3$ and
$\phi^4$ which are multiplied by dimensionless couplings. However
we will not include these terms in the lagrangian
\footnote{The terms with odd powers of $\Phi$ can be removed by
imposing symmetries of the strong interaction, while $\phi^4$ is
assumed to be absent through a proper choice of the form factors
in the strong coupling regime.}. The symmetry breaking Lagrangian
${\cal L_{SB}}$ is supposed to contain scale-non-invariant terms,
in particular, a stabilizing (stringy) potential for $\phi$ in the
S-frame. For this reason we write: \beq {\cal L}_{\rm
SB}=-\sqrt{-g} (a \phi^2 + b + c \frac{1}{\phi^2}). \label{SB}
\eeq

Happily, it is possible to satisfy the field equations with
constant values of the fields $\phi$ and $\Phi$ through a proper
choice of positive (but not fine-tuned) values of the parameters
$a, b, c$, maintaining $f<0$ and $\lambda_{\Phi}>0$. We made sure
that $g_s>1$ can be recovered in the equilibrium configuration and
that, consequently, the solution is consistent with the
non-perturbative action that we considered as a starting point.

Here is a possible choice of parameters (in string units):
$f=-4$, $\lambda_{\Phi}=27$, $a=\frac{1}{9}$, $b=c=\frac{1}{72}$. In the equilibrium configuration we have
$\phi_0=\frac{1}{2}$ and $\Phi_0=\frac{1}{3}$.

As we will see, the description in the E-frame will guarantee the
presence of scale-invariance on cosmological distances. This will
be the crucial element in our analysis to address the cosmological
constant problem. Remarkably, even if we stabilize the dilaton in
the S-frame, the conformal transformation to the E-frame will be
non-trivial. This point needs to be further elaborated. Let us
consider a stabilizing potential $V(\sigma)$ for the dilaton in
the S-frame and let us call $\sigma_0$ the value of the dilaton in
the minimum of the potential. When we perform the conformal
transformation, the minimum of the potential $V(\sigma_0)$ will be
multiplied by the conformal factor $e^{-4 \zeta \sigma_0}$ (which
is constant). A different point of the potential, for example
$V(\sigma_1)$, will be multiplied by a different constant
conformal factor (i.e. $e^{-4 \zeta \sigma_1}$). Consequently, the
function $V(\sigma)$ will be multiplied by a non-constant function
of the dilaton, namely, a non-trivial conformal factor given by $\xi^{-2} \phi^{-4}=e^{-4 \zeta \sigma}$.
Therefore, the potential \ref{SB} will be mapped by the conformal transformation into an E-frame potential given by
\beq V=\xi^{-2}[a \phi^{-2}+b \phi^{-4}+c \phi^{-6}]= a e^{-2\zeta \sigma}+b e^{-4\zeta \sigma}+c e^{-6\zeta\sigma}. \label{RA}\eeq
Remarkably, the parameters $a$, $b$ and $c$ are {\it positive} and, consequently, the potential \ref{RA} is run-away towards the region of weak coupling.
This is a first hint that different dynamical behaviours of the
dilaton in different frames are not forbidden. We warn the reader
that the ``dictionary" of the conformal transformation is still
valid (i.e. we still write $\phi= \xi^{-1/2} e^{\zeta \sigma}$),
but, as we will see in the following sections, in this model a
stabilized dilaton in one frame (S-frame) does {\it not}
correspond to a stabilized dilaton in another frame (E-frame). At
first glance this result seems to clash with the approach
discussed in \cite{Zanzi:2006xr}, therefore, we will try to
overcome this problem in a future work.

\section{The scale invariant Lagrangian}
\label{SI}

In this section we will analyze ${\cal L}_{\rm SI}$ following
\cite{Fujii:2002sb, Fujii:2003pa}.

\subsection{Classical level}

\subsubsection{Jordan frame}

${\cal L}_{\rm SI}$ is attractive because the coupling constant
$f$ is {\em dimensionless}, hence vesting scale invariance in the
gravity-matter system. By applying Noether's procedure we obtain
the dilatation current as given by \beq J^{\mu}=\frac{1}{2}
\sqrt{-g}g^{\mu\nu} \partial_{\nu} ( \xi \zeta^{-2} \phi^2 +
\Phi^2), \label{cfr1-73b} \eeq which is shown to be conserved by
using the field equations.

We summarize the main features of the scale-invariant part of the
model in the Jordan frame (J-frame) as follows:
\begin{itemize} \item the model is scale invariant; \item $\Phi$
is massless; \item there is a direct coupling between matter and
scalar field.
\end{itemize}

\subsubsection{Einstein frame}
\label{vacuum}

Since our intention is to take into account quantum effects, {\it
regularization} will be necessary and whatever will be the
regularization method we will choose, a scale will be introduced
in the theory leading to a breaking of scale invariance. In the
following we will consider the method of continuous dimensions
and, for this reason, we start writing conformal transformation
equations in D-dimensions. To proceed further, the lagrangian
\reflef{bsl1-96}) will be rewritten in the E frame and {\it
spontaneous} breaking of scale invariance will be discovered.

In D=2d dimensional spacetime a scalar field has canonical
dimension (d-1) and the conformal transformation looks like
\begin{equation}
\sqrt{-g}=\Omega^{-D} \sqrt{-g_*}. \label{ctdd}
\end{equation}
In order to recover the usual Einstein-Hilbert term we impose
\begin{equation}
\xi \phi^2=\Omega^{D-2} \label{ctdd}
\end{equation}
that implies
\begin{equation}
\Omega=e^{\frac{\zeta \sigma}{d-1}}, \label{ctdd}
\end{equation}
where the relation \reflef{bsl1-60}) and $M_p=1$ have been
exploited.

In this way we can rewrite the lagrangian \reflef{bsl1-96}) in the
E frame as
\begin{equation}
{\cal L}_{*}=\sqrt{-g_*}\left( \frac{1}{2} R_* -
    \half g^{\mu\nu}_*\partial_{\mu}\sigma\partial_{\nu}\sigma + {\cal L}_{* matter}
    \right),
\label{eframe}
\end{equation}
where ${\cal L}_{* matter}$ turns out to be
\begin{equation}
{\cal L}_{* matter}= -
    \half g^{\mu\nu}_* D_{\mu}\Phi_* D_{\nu} \Phi_* - e^{2 \frac{d-2}{d-1} \zeta
    \sigma} (\xi^{-1} \frac{f}{4} M_p^2 \Phi_*^2+ \frac{\lambda_{\Phi}}{4!} \Phi_*^4)
\label{lmatter}
\end{equation}
and $D_{\mu}=\partial_{\mu}+ \zeta \partial_{\mu} \sigma$.

The conservation law remains true even after the conformal
transformation, but the symmetry is broken {\em spontaneously} due
to the trick by which a dimensionful constant $M_{\rm P}(=1)$ has
been "re-installed" in \reflef{lmatter}). The dilatation current
in the E-frame takes the form:
\begin{equation}
J^{\mu}=\frac{1}{2} \sqrt{-g_*} g_*^{\mu \nu} [2 \zeta^{-1}
\partial_{\nu} \sigma + (\partial_{\nu} + 2 \zeta (\partial_{\nu}\sigma) ) \Phi_*^2
],
\label{correnteEframe}
\end{equation}
while the generator
\begin{equation}
Q=\int d^3 x J^0 \label{generatore}
\end{equation}
does {\it not} annihilate the vacuum due to the presence of the
term linear in $\sigma$ in (\ref{correnteEframe}). In other words,
scale invariance is {\it spontaneously} broken and we write
\begin{equation}
Q|0\rangle \neq 0.
\end{equation}
In this context $\sigma$ is a Nambu-Goldstone boson, a dilaton.

In the classical theory, at this stage, we can put $D=4$ or $d=2$
in \ref{lmatter} and we obtain a dilaton $\sigma$ coupled to
matter only through the derivative coupling included in $D_{\mu}$.
This means that at the classical level, at least in the static and
long-wavelength limit, we have no way to observe the dilaton by
measuring forces between matter objects. Moreover, if we focus our
attention on the last two terms in \ref{lmatter}, putting $d=2$
will give us a Higgs-like lagrangian. For this reason we now apply
the same recipe as with the Higgs sector in the Standard Model, we
split $\Phi_*$ into a vacuum expectation value (VEV) and a fluctuating field in the form
\begin{equation}
\Phi_*=v_{\Phi} + \tilde\Phi \label{splitting}
\end{equation}
and we rewrite the last two terms in \ref{lmatter} as
\begin{equation} -L'_0=L_{vac} + \frac{1}{2} m^2 \tilde\Phi^2 +
\frac{1}{2} \sqrt{\frac{\lambda_{\Phi}}{3}} m \tilde \Phi^3 +
\frac{\lambda_{\Phi}}{4!} \tilde \Phi^4, \label{higgslike}
\end{equation}
where $m^2=-\frac{f}{\xi} M_p^2$ and
\begin{equation}
L_{vac}= - \frac{3}{8} \frac{f^2}{\xi^2 \lambda_{\Phi}} M_p^4
\varpropto v_{\Phi}^4 \label{cc}
\end{equation}
is a vacuum energy term.

It is noteworthy to summarize the main features of ${\cal L_{SI}}$
in the E-frame at the {\it classical} level:\begin{itemize}\item
$\Phi_*$ has a constant mass; \item no potential is present for the
dilaton, which remains strictly massless, one more signal that the
breaking of scale invariance is only {\it spontaneous} at the
classical level. However, we will deal with {\it explicit}
breaking of scale invariance when we will quantize the theory in
the following paragraphs: in that case we will also find a
non-vanishing dilatonic potential (i.e. dilatonic mass). \item At
the classical level the dilaton disappears for d=2 and the only
direct coupling matter-dilaton is included in the derivative term
$D_{\mu}$.
\end{itemize}
\subsection{Quantum level}
In this section a direct coupling between matter and dilaton will
re-emerge as the result of a quantum anomaly. In particular we are
going to touch upon these points: 1) quantum effects and conformal
anomaly; 2) explicit breaking of scale invariance and
Pseudo-NG-boson nature of the dilaton.

\subsubsection{Anomaly and dilaton coupling}

If we substitute \reflef{splitting}) and the expansion
\begin{equation}
e^{2 \frac{d-2}{d-1} \zeta
    \sigma}=1+ 2 \zeta \frac{d-2}{d-1} \sigma +... \label{espansione}
\end{equation}
in formula \reflef{lmatter}), we obtain the following lagrangian
at first order in $\sigma$:
\begin{equation}
-{\cal L}_{1}= 2 \zeta (d-2) \sigma \left( \frac{1}{2} m^2 \tilde
\Phi^2 + \frac{1}{2} \sqrt{\frac{\lambda_{\Phi}}{3}} m \tilde
\Phi^3 + \frac{\lambda_{\Phi}}{4!}  \tilde \Phi^4 \right) -
\frac{3}{4} \zeta (d-2) \sigma \frac{f^2}{\xi^2 \lambda_{\Phi}}
M_p^4 . \label{interazioni}
\end{equation}

In this way several different 1-loop diagrams can be constructed
(Figure 1): we will consider an explicit calculation starting from
diagram (c).

\begin{minipage}[t]{14.4cm}
\baselineskip=0.4em \epsfxsize=14cm
\mbox{}\\[-7.4em]
\hspace*{-2.em} \epsffile{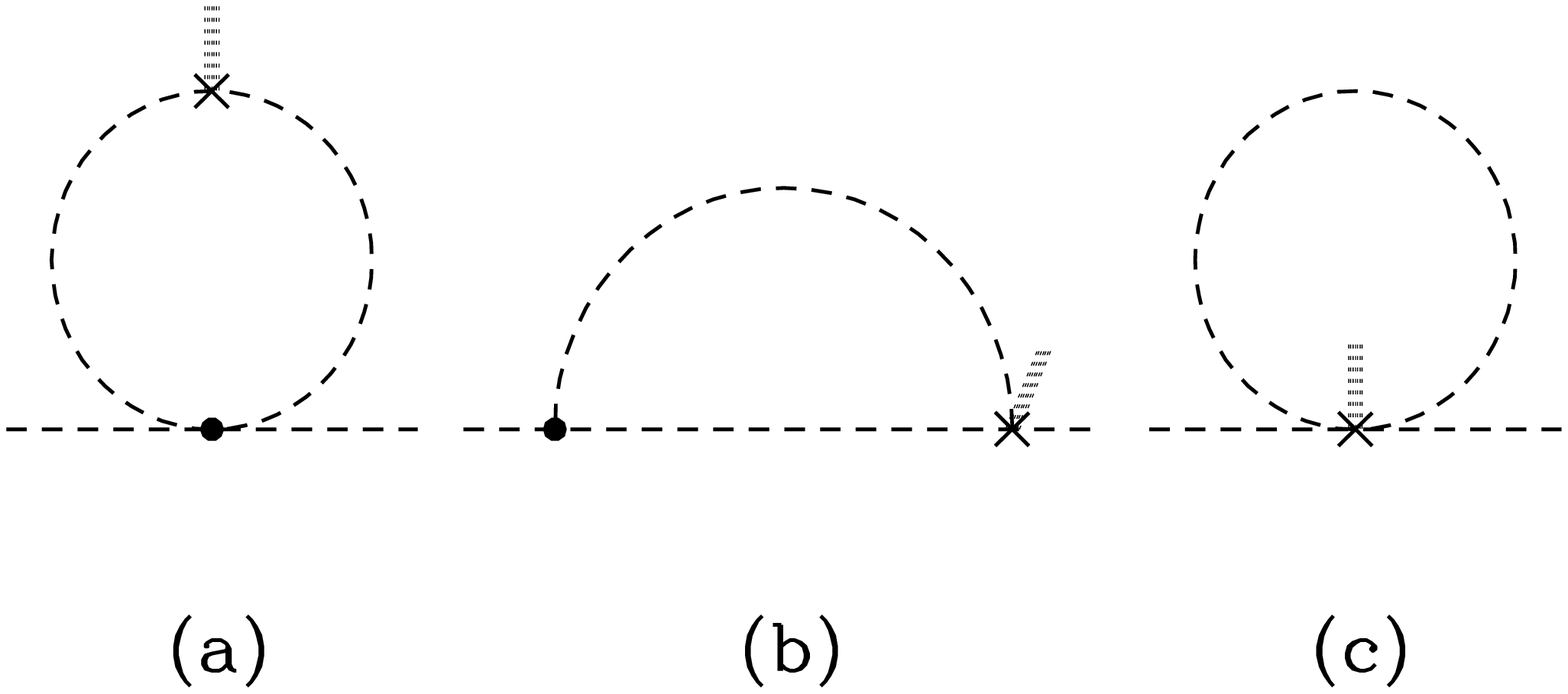}
\mbox{}\\[-5.1em]
Figure 1.  {\sl Examples of 1-loop diagrams for the interaction
$\lmd_\Phi \Phi_*^4$, represented, together with the derived
3-vertex, by a filled circle.
Heavy dotted lines are for $\sigma$. This figure can be found in reference \cite{Fujii:2004bg}.}  \end{minipage}
\label{diagrammi}
\mbox{}\\[-.0 em]

The one-loop diagram is divergent. Exploiting dimensional
regularization the contribution given by the scalar loop is \beq
{\cal I}_{c}=-i \zeta \lambda_{\Phi} (D-4) (2 \pi)^{-D} \int
d^{D}k \frac{1}{\left( k^2 + m^2 \right)}, \label{intloop} \eeq
where the integral in D-dimensions is evaluated explicitly as \beq
I_{c}= \int d^{D}k \frac{1}{\left( k^2 + m^2 \right)} = i \pi^2
(m^2)^{d-1} \Gamma(1-d). \label{intloop} \eeq

Remarkably (2-d) cancels out the pole at d=2 in $\Gamma(1-d)$
leading to a finite result in 4 dimensions:

 \beqa (2-d )\Gamma(1-d )&=&\frac{1}{1-d}(2-d
)\Gamma(2-d )
\nnb\\
&=& \frac{1}{1-d}\Gamma(3-d )\stackrel{d \rightarrow 2}
{\longrightarrow} -1\label{cfrc_4} \eeqa and this is one of the
possible ways in which an anomaly may present itself.

Adding up the contributions from the three diagrams in Figure 1 we
obtain \beqa {\cal I}_{tot}={\cal I}_a + {\cal I}_b+{\cal I}_c=
\frac{1}{\pi^2} \zeta \lambda_{\Phi} m^2\label{vertice} \eeqa or,
in other words, the fundamental (anomalous) interaction vertex
between dilaton and matter \cite{Fujii:2002sb}: \beqa {\cal
L}_{\Phi \Phi \sigma}=-\frac{1}{2} \frac{{\cal I}_{tot}}{M_p}
\tilde \Phi^2 \sigma.\label{vertice} \eeqa

There are a number of consequences of this last formula. Here we
mention: \begin{itemize} \item weak equivalence principle (WEP) is violated.

\item Scale invariance is broken explicitly and the dilatation
current is not divergenceless anymore. In particular we can write
$\partial_{\mu} J^{\mu} = \zeta^{-1} \mu_{\sigma}^2 \sigma + ...$.

\item According to the relativistic quantum field theory, the
"anomaly-induced" interaction \reflef{vertice}) with the matter
field $\Phi$ leads us to the contributions depicted in Figure 2.
Naturally every diagram will give a (too) large contribution to
the unrenormalized vacuum energy. In the proposal of Fujii
\cite{Fujii:2002sb}, the difference between the (dilatonic) unrenormalized
vacuum energy and the cosmological constant scale is due to the
dilatonic backreaction. In other words, he split the dilaton in
two components: $\sigma=\sigma_b(t)+\sigma_f(x)$ and the effects
of $\sigma_f$ on $\sigma_b$ are supposed to renormalize the vacuum
energy and to render it compatible with the Dark Energy scale on
cosmological distances. We will further elaborate on these issues
in the following sections. It seems worthwhile to point out that
the set of diagrams in figure 2 must be considered as an expansion
to {\it all orders} in perturbation theory, namely as a complete
expansion and not as a 1-loop contribution. In other words, since
the interaction vertex between dilaton and matter has been
obtained from a 1-loop calculation (from the conformal anomaly),
every time we add one external leg we are taking into account an
additional loop in the calculation.
\end{itemize}

\begin{minipage}[t]{14.4cm}
\baselineskip=1.7em \epsfxsize=12cm
\mbox{}\\[-0.4em]
\hspace*{-0.1em} \epsffile{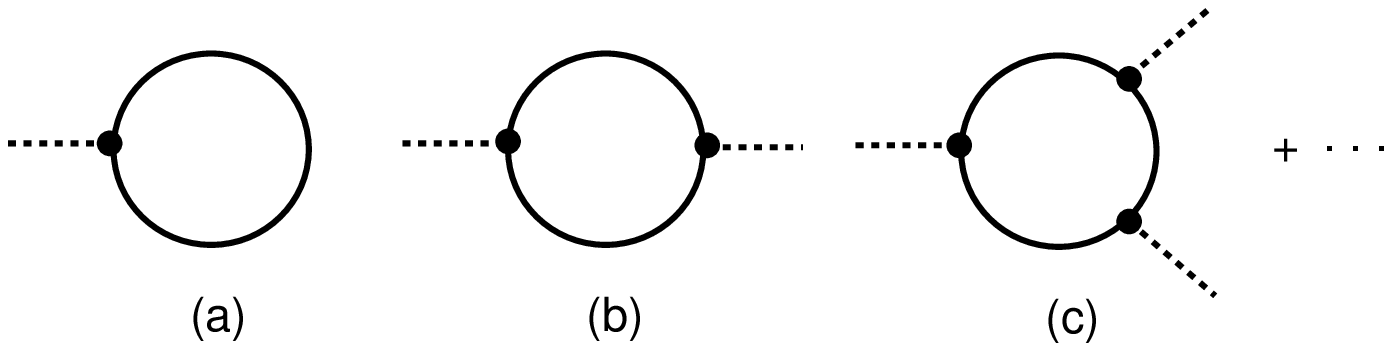}
\mbox{}\\[-0.1em]
Figure 2.  {\sl First three examples of the 1-loop diagrams for
the quantum corrections due to the "anomaly-induced" matter
coupling. Solid and dotted lines are for
$\Phi$ and $\sigma$ respectively. This figure can be found in reference \cite{Fujii:2002sb}.}  \end{minipage}
\mbox{}\\[-.0 em]

\setcounter{equation}{0}
\section{A stringy solution to the cosmological constant problem}
\label{CC}

As already mentioned in the Introduction, several points are
unsatisfactory about the proposal of Fujii in reference
\cite{Fujii:2002sb}, namely:
\begin{itemize}
\item the theoretical origin of the model is not discussed.

 \item
A globally massless but locally massive dilaton can be accepted
and it is not forbidden by the renormalization program. It would
be rewarding to show that (1) this result {\it must} be obtained
and (2) it admits a simple description.

\item It would be rewarding to connect the dilatonic
backreaction mentioned by Fujii with the cosmological (i.e. metric) backreaction.

\item A detailed mechanism
to suppress the contribution to the vacuum energy coming from
quantum diagrams with dilatons in the external legs is missing. In other words, in the proposal of Fujii
we know that the dilatonic backreaction exists, but we do {\it not} know whether it is effective as counterterm for the dilatonic vacuum energy.
Moreover, Fujii carried on his
calculations only for the diagrams with one and two external dilatons: the
remaining diagrams (present of course in infinite number) have not
been calculated.
\item The contribution to the vacuum energy coming from matter fields,
gauge fields and gravitons is not discussed.

\item The origin of the correct DE scale is
unknown;
\end{itemize}
In this section we will overcome all these problems. The final
outcome is a totally new (and stringy) solution to the
cosmological constant problem.

About the organization of this section, we will proceed stepwise
solving all the problems mentioned above (naturally we have
already pointed out the stringy origin of our model in section
\ref{modello}):\\
1) in subsection \ref{theorigin} we suggest that a conceivable interpretation of the dual mass
of the dilaton is simply a chameleonic behaviour of the field.
This will be our {\it final result} in the effective potential of
the theory in the E-frame and it provides a simple description of
the dual mass.
Remarkably, this result {\it must} be obtained in our approach.\\
2) In subsection \ref{thenature} we establish a connection between the dilatonic and the metric backreactions.\\
3) In subsection \ref{symmetryrestoration} we show that scale-invariance is almost restored on
cosmological distances (in the Einstein frame). In this way on
cosmological distances we protect, on the one hand, the mass of
the dilaton, on the other hand, the cosmological constant.
Our argument is valid at all orders in perturbation theory. Naturally this symmetry principle guarantees that the dilatonic backreaction {\it must} suppress the dilatonic vacuum energy (i.e. the renormalization of the dilatonic vacuum energy must be successful). Moreover, we suggest in this article to exploit the metric backreaction as counterterm in the renormalization of the gravitational vacuum energy. Once again, our symmetry principle guarantees that the metric backreaction {\it must} suppress the gravitational vacuum energy. As far as matter and gauge fields are concerned, we will show that their contribution to the cosmological constant is properly suppressed.\\
4) In subsection \ref{thecorrect} the correct DE scale is
recovered in the E-frame without unnatural fine-tunings. However,
at this stage, the possibility still exists that a certain amount
of fine-tuning may be required to satisfy some phenomenological
constraints. A detailed phenomenological analysis is definitely
necessary to clarify this point.\\

\subsection{The origin of the dual mass}
\label{theorigin}

We suggest that a conceivable interpretation of the dual mass of
the dilaton $\sigma$ is simply a chameleonic behavior of the
field. This will be our {\it final result} at the level of the
effective potential in the E-frame and it provides a simple
description of the dual mass. As we will see, in our approach this
result {\it must} be present.

\subsection{The nature of the backreaction}
\label{thenature}

In the Fujii's proposal, the backreaction must be considered as
effects from $\sigma_f$ on $\sigma_b$. We now connect this
dilatonic backreaction with the cosmological
backreaction recently discussed in the literature (see
\cite{Rasanen:2003fy, Kolb:2004am, Kolb:2005da}).

\subsubsection{The metric backreaction}

In the proposal of \cite{Rasanen:2003fy, Kolb:2004am, Kolb:2005da}
the backreaction of properly smoothed-out inhomogeneities can
account for an accelerated expansion of the Universe. If our
intention is to describe the local inhomogeneities, we can start
considering the Einstein equations for the metric. Since the
equations are non-linear, typically the averaged equation will not
correspond to the equation for the averaged metric. In more
detail, if we consider a metric $g_{\alpha \beta}$ and we
construct the Einstein tensor $G_{\mu\nu}(g_{\alpha\beta})$, in
general we have: \beq <G_{\mu\nu}(g_{\alpha\beta})> \neq
G_{\mu\nu}(<g_{\alpha\beta}>).\label{vale}\eeq This difference
between the Einstein tensor obtained from the smoothed-out metric
and the smoothed out matter tensor represents the effect (i.e. the
backreaction) of small-scale inhomogeneities on the dynamic at the
smoothed-out scale. For a review see for example
\cite{Marra:2008sy}.

\subsubsection{The dilatonic backreaction}

Let us come back to the dilaton. Local
inhomogeneities in the matter density will introduce a length
scale into the problem and scale invariance will be broken. We
point out that these fluctuations of the matter density are
directly connected with the fluctuations of a chameleon: the
minimum of the effective potential of a chameleon depends on the
density of the environment. Consequently, the fluctuations of a chameleon will define the length scale for the averaging procedure.
In this way we suggest a connection between the chameleonic (dilatonic)
backreaction and the cosmological (i.e. metric) one of
references \cite{Rasanen:2003fy, Kolb:2004am, Kolb:2005da}.

Let us consider now a diagram with N external legs in figure 2.
This diagram is a function $F_N(\sigma)$ of the dilaton.
Expectation values of quantum operators can be rewritten as
spatial integrals weighted by the integration volume (see for
example \cite{Gasperini:2009wp}). Following the same procedure of
eq. \ref{vale}, whatever will be the diagram of figure 2 we
consider, we can write: \beq <F_N(\sigma)> \neq F_N(<\sigma>).
\label{BRrin}\eeq In this last equation $<F_N(\sigma)>$ is the
contribution of the diagram to the unrenormalized vacuum energy
(and it is extremely large), while $F_N(<\sigma>)$ enters directly
in the effective potential for the dilaton, it takes into account
the dilatonic backreaction and, in the proposal of Fujii, it is
not forbidden to think that it is small on cosmological distances.
Happily, in the next subsections we will point out that this turns
out to be the case.

\subsection{Symmetry restoration}
\label{symmetryrestoration}

We now suggest that scale invariance is almost restored on
cosmological distances in the E-frame and the dilaton
in the E-frame is parametrizing the amount of scale invariance of the problem.

To illustrate this point, let us mention once again that the
quantum dilatonic potential of formula \ref{SB} becomes run-away
after the conformal transformation to the E-frame, because the
conformal factor introduces an exponential suppression. Naturally,
the larger is the value of $\sigma$, the more effective will be
the suppression. Therefore, the question is: {\it what is the VEV
of $\sigma$?} The standard approach to find this VEV would be to
evaluate the effective action of the model exploiting the
deWitt-Vilkovisky method (for an introduction on this subject see
for example \cite{Buchbinder:1992rb}). However, in the absence of a detailed analysis,
but inspired by chameleon theories, we
parametrize the length scales of the problem through the value of
$\sigma$. In this way, we can consider $\sigma_b$ and $\sigma_f$
as two distinct (but related) objects corresponding to different
length scales (see also section \ref{azionieffettive} below). Accordingly, we write $\sigma_b \neq \sigma_f$. It
follows that a constant term in the J-frame, which explicitly
violates scale invariance on {\it all} distances, will be
exponentially reduced in the E-frame, but only for large values of
$\sigma$. In other words, even if our starting point is a J-frame
theory which is {\it not} scale-invariant, a constant term in the
J-frame will be suppressed for large $\sigma$ after the conformal
transformation to the E-frame.

Let us discuss this issue for the various {\it fields}:
\begin{itemize}
\item {\it The spontaneous breakdown of scale-invariance
introduced by the conformal transformation to the E-frame is
negligible for large $\sigma$.}

As far as the matter sector is concerned, $\Phi$ has finite constant value $\Phi_J$ in the J-frame and the conformal transformation introduces the exponential
rescaling of the $\Phi_*$-field.

As far as the Einstein-Hilbert term is concerned, it is not scale
invariant and we will now show that it is (almost) negligible for
large $\sigma$. This result is based on two elements, namely: (1)
the action \ref{Ltotale} in the S-frame is the result of a {\it
non-perturbative} calculation and (2) quantum
loops can induce a kinetic term for gravitons (i.e. induced
gravity mechanism \cite{Sakharov:1967pk}), see figure 3.
Accordingly, we suggest to induce a S-frame kinetic term for gravitons through quantum diagrams. Therefore, the Einstein-Hilbert term in formula \ref{eframe}
is coming from quantum diagrams. However, these diagrams are
suppressed in the E-frame for large $\sigma$,  because {\it all}
interactions are switched off for large $\sigma$
\cite{Witten:1984dg} (i.e. in the weak coupling regime of String
theory). Consequently, the larger is $\sigma$ in the E-frame, the
smaller is the Einstein-Hilbert term. We will further elaborate on
this non-perturbative induced gravity idea in the following
subsections.

\begin{minipage}[t]{14.4cm}
\baselineskip=1.7em \epsfxsize=12cm
\mbox{}\\[-0.4em]
\hspace*{-0.1em} \epsffile{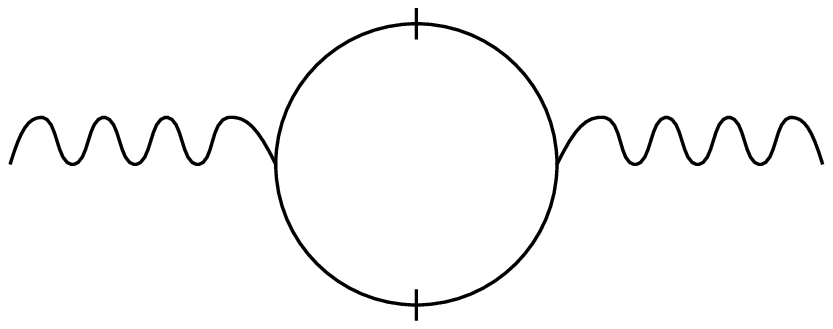}
\mbox{}\\[-0.1em]
Figure 3.  {\sl The  one-loop diagram generating the
Ricci scalar for the gravitons. Wave lines denote gravitons,
solid lines denote massive scalars/fermions. Vertical short lines on
scalar/fermion propagators indicate that they are massive. This figure can be found in reference \cite{Dvali:2000hr}.}  \end{minipage}
\mbox{}\\[-.0 em]

\item {\it For large $\sigma$, conformal
anomaly is harmless and all the interactions are switched off.} A
scale-dependence of the conformal anomaly has already been
discussed in the literature. For example, in the framework of
AdS/CFT correspondence \cite{Maldacena:1997re} (for a review see
for example \cite{Aharony:1999ti, Maldacena:2003nj}), an
increasing central charge as a function of the energy scale has
already been discussed (the reader is referred to
\cite{Nojiri:1999eg} and references therein). To the best of our
knowledge, this should be a rather general property of conformal
anomalies in agreement with c-theorems (for a review see for
example \cite{Ogushi:2001md}). In our model, this result is
obtained through the exponential rescaling mentioned above. In
more detail, we can write the anomalous coupling \ref{vertice} as:
\beq \frac{1}{2}{\cal I}_{tot} \tilde{\Phi}^2 \sigma =
\frac{1}{2}{\cal I}_{tot} e^{-2 \zeta \sigma} \Phi_f^2 \sigma,
\label{anomaliaexp} \eeq where $\Phi_f$ is the fluctuating
component of $\Phi$ in the J-frame and it is constant. We infer
that for large values of $\sigma$ the coupling is suppressed.

A harmless conformal anomaly for large $\sigma$ will be a crucial element in our analysis, where, as we will show below, large values of $\sigma$ will correspond to large distance scales.
A word of caution is necessary: conformal anomaly is not always harmless in Cosmology.
Indeed, the trace of the energy momentum tensor enters in the chameleonic equations and it can influence the dynamical behaviour of the field. For relativistic degrees of freedom it is generally assumed that this trace is vanishing: not so at all. The trace receives a contribution, on the one hand, when a particle species becomes non-relativistic, on the other hand, from trace anomaly. In particular, it was shown (see for example \cite{Brax:2004qh}) that the effective equation of state for a plasma of an $SU(N_c)$ group with coupling $g$ and $N_f$ flavours is given by:
\beq 1-3w=\frac{5}{6 \pi^2} (\frac{g^2}{4 \pi})^2 \frac{(N_c + \frac{5}{4} N_f)(\frac{11}{3} N_c - \frac{2}{3} N_f)}{2+ \frac{7}{2} N_c N_f/(N_c^2 -1)}+ {\cal O}(g^5). \label{anomaliasun}\eeq
Remarkably, if we switch off the interaction in this last formula, we recover $w=1/3$ and trace anomaly is globally harmless.
In our stringy model, since the dilaton controls the strength of the interactions, {\it all} the interactions ({\it including gravity}) will be switched off for large $\sigma$ (i.e. weak coupling regime). Indeed, string theories imply a natural unification of the
couplings: gauge and gravitational couplings in heterotic string
theory always automatically unify at tree-level to form the string
coupling constant $g_s$ (which is determined by the dilaton). Consequently, we are considering a {\it free} string theory when $\sigma$ is large. For this reason,
in the remaining part of the paper we will {\it always} neglect conformal anomaly for large $\sigma$.

\end{itemize}

We infer that for large $\sigma$ scale invariance is restored in the E-frame.
Remarkably, there are a number of consequences of this
restoration of scale-invariance:\\

{\it Consequence 1}.  $\sigma_b$ and $\sigma_f$ acquire different
mass because the degree of divergence of their loop diagrams is
different.\\

{\it Consequence 2}. We extend consequence 1 to all the diagrams
(present of course in infinite number) shown in figure 2.\\

{\it Consequence 3}. For large $\sigma$ the (almost) symmetric
configuration will be compatible neither with large couplings between dilaton and matter, nor
with large mass scales, including the cosmological constant and the mass of the dilaton. Therefore, the total renormalized vacuum energy
is run-away because scale invariance is restored for large $\sigma$.\\

From these three consequences, since we know that $\sigma$ gets a mass through
the interaction with matter (encoded in formula \ref{vertice}), we infer
that $\sigma_b>\sigma_f$ and, therefore, scale invariance is restored on cosmological distances.
In more detail, the mass of the dilaton is obtained through a competition between the run-away branch of the potential and the matter branch. Therefore, the total unrenormalized vacuum energy must be {\it positive}. Had we considered a negative (total) unrenormalized vacuum energy, there would have been no competition between the run-away branch of the potential and the matter branch and this would have clashed with consequences 1 and 2.

Naturally we have an infinite number of different contributions to
the cosmological constant. Needless to say, a vacuum energy term
on short distances does {\it not} contribute to the Dark Energy.
Therefore, the contributions to the vacuum energy which are
relevant for the cosmological constant are {\it global} and not
local. This last comment requires a more detailed discussion.
Expectation values of quantum operators can be rewritten as
spatial integrals weighted by the integration volume (see for
example \cite{Gasperini:2009wp}). On the one hand, if we consider
an integration volume much smaller than the volume of the
Universe, we are dealing with a {\it local} quantity and the
result will be the expectation value of the field (EV).  On the
other hand, if we consider the spatial integral of the field
weighted by the volume of the visible Universe, we are
dealing with a {\it global} (cosmological) object and the result
is the {\it vacuum} expectation value of the field (VEV), which is
the relevant one in the evaluation of the Feynman diagrams
contributing to the cosmological constant. In this way, our
solution to the cosmological constant problem is intrinsically
linked to a {\it dual nature} of the concept of particle. The splitting of a particle in a background (global) part
and in a fluctuating (local) part, that we already discussed for
the dilaton, is extended in our proposal to {\it all the fields}.
Whatever will be the Feynman diagram we consider, if our intention is to evaluate its contribution to the cosmological constant (i.e. to the global renormalized vacuum energy), we must (1) construct the diagram exploiting only the background part of the particles and (2) give a VEV to the external legs (if they are present).
An interesting line of development will elaborate on
this dual nature of the concept of particle starting from
\cite{Colosi:2004vw, Buchbinder:1992rb, DeWitt:2003pm} and references therein. We discuss separately
the different particles and their contribution to the cosmological constant
in the following (sub)sections. This will require the analysis of three different categories of (global) Feynman diagrams: A) the diagrams with the particle in the external legs; B) the diagrams with the particle in the internal legs; C) bubble diagrams.

\subsubsection{The dilaton}

{\it The dilaton in the external legs.} Since the local dilatonic vacuum energy is very different from the DE scale, when we average the local dilatonic vacuum energies on large distances,
an unacceptably large contribution is expected. This is the cosmological dilatonic vacuum energy {\it before} the renormalization and it does {\it not} correspond to the DE. In other words, if we construct a diagram of figure 2 exploiting only the background part of the fields, a very large (unrenormalized) vacuum energy will be obtained if we don't give a VEV to the external dilatons. As already pointed out by Fujii, the difference between the unrenormalized dilatonic vacuum energy and the DE scale is due to (dilatonic) backreaction effects. The global {\it renormalized} vacuum energy is obtained {\it by giving a VEV} to the background external particles ($\sigma_b$ in this case), see also formula \ref{BRrin}. We claim that this
strategy {\it must} be successful, because we know that scale invariance is restored for large $\sigma$.\\

{\it The dilaton in the internal legs.} It seems noteworthy that scale invariance will forbid a
dilatonic force on cosmological distances. The interaction dilaton-matter is encoded in the conformal anomaly and it is globally harmless. We infer that the global dilaton is harmless in the internal legs. One last remark is in order. If we
remember that on short distances the kinetic term of $\sigma_b$ is
negligible with respect to the kinetic term of $\sigma_f$, we can
conclude that $\sigma_b$ does not mediate a force between
macroscopic objects. Consequently, as already pointed out by
Fujii, only $\sigma_f$ mediates a dilatonic force.\\

{\it The 1-loop bubble diagram with a dilaton.} The careful reader may be worried by the presence of bubble diagrams (i.e. Feynman diagrams without external legs). In particular, we can construct a 1-loop bubble diagram with the dilaton field. Once again, the diagram with $\sigma_f$ will give a large contribution to the vacuum energy, but the relevant diagram for the evaluation of the dilatonic contribution to the cosmological constant is constructed with $\sigma_b$. What is its contribution? We point out that the bubble diagram represents a quantum breakdown of scale-invariance. In other words, it is a conformally anomalous contribution and, for this reason, it will be negligible on cosmological distances. This last point requires a more detailed discussion. In the plots of figure 2, the interaction vertex between dilaton and matter is the result of a conformal anomaly and it is globally harmless because the vertex is exponentially suppressed on large distances (see formula \ref{anomaliaexp}). It must be stressed that in the simplest bubble diagram this argument cannot be applied because there are no vertices. However, since the dilaton is running towards the weak coupling regime, all the interactions are switched off on cosmological distances and, therefore, we can neglect the contribution of the bubble diagram because the related conformal anomaly will be globally harmless (see the discussion about formula \ref{anomaliasun}). We infer that the restoration of scale invariance on cosmological distances is not affected by the 1-loop dilatonic bubble and the cosmological constant remains under control.\\

\subsubsection{Matter particles (spin-0, spin-1/2) and Rarita-Schwinger spin-3/2 fields}

{\it Matter particles in the external legs.} The VEV of the matter fields is almost vanishing, because we can write $\Phi_*=\Phi_J e^{- \zeta \sigma}$.\\

{\it Matter particles in the internal legs.} In the case of internal legs, we don't give a VEV to the fields. The matter contribution is still negligible, because global matter is not interacting.\\

{\it Bubble diagrams with matter.} Matter bubble diagrams give conformally anomalous contributions and, consequently, they are globally harmless.\\

Therefore, the quantum contributions to the global renormalized
vacuum energy are under control. For this reason, the vacuum
energy of matter fields discussed in section \ref{vacuum}, namely
$L_{vac}=-\frac{3}{8} \frac{f^2}{\xi^2 \lambda} M_p^4$, can be
safely neglected for large $\sigma$.

The contribution of spin-3/2 particles to the cosmological constant can be analyzed following the procedure discussed for matter particles.

\subsubsection{Gauge particles: spin-1}

{\it Gauge particles in the external legs.} Even if spin-1
gauge fields are not affected by the conformal transformation
($A_{\mu}=A_{\mu}^*$, see for example \cite{Fujii:2007qv}), this is not a problem, because the presence of unbroken
fundamental symmetries (e.g. $SU(3)_C \times U(1)_{em}$,
Lorentz...) requires a vanishing VEV for the gauge fields and, therefore, external global gauge particles
are harmless.\\

{\it Gauge particles in the internal legs.} Gauge interactions are switched off globally and, therefore, internal (global) gauge particles are harmless.\\

{\it 1-loop bubble diagram with a gauge field.} Once again, this contribution is a conformal anomaly and, consequently, it is globally harmless.

\subsubsection{The graviton}

Let us start writing the complete E-frame metric as:
\beq g^{\mu\nu}_*=(g^{\mu\nu}_*)_b + (g_*^{\mu\nu})_f.\eeq
Let us consider the two metrics separately.

The local metric will produce a (large)
local gravitational vacuum energy. If we divide the visible Universe into small bubbles, whatever will be the bubble we choose, the local gravitational vacuum energy will be much larger than the DE scale. We infer that the average value (over the volume of the visible Universe) of all these local gravitational vacuum energies inside the bubbles will be much larger than the meV scale (this is the same argument we discussed for the dilatonic backreaction). However, this is the {\it unrenormalized} global gravitational vacuum energy and it does not correspond to the DE. We suggest in this paper to exploit the metric backreaction as a counterterm in the renormalization process of the gravitational vacuum energy. In other words, we extend the proposal of Fujii from the dilatonic to the metric vacuum energy. We also claim that the metric backreaction will be able to properly suppress the cosmological gravitational vacuum energy, because scale-invariance is almost restored on large distances.

As far as the global (cosmological) metric is concerned, we choose $(g^{\mu\nu}_*)_b=g^{\mu\nu}_{FRW}$, where $g^{\mu\nu}_{FRW}$ is the usual Friedmann-Robertson-Walker (FRW) metric. As already mentioned above, $\sigma_b$ does not mediate a force between macroscopic objects. The same result is obtained in the case of the FRW-metric, because, in our approach the kinetic term for the gravitons is induced by quantum diagrams. In this way, since in our proposal {\it all} the interactions are switched off on cosmological length scales in the E-frame, the kinetic term for the global gravitons will be subdominant with respect to the kinetic term for the local gravitons. Therefore, the global metric $(g^{\mu \nu}_*)_b$, like $\sigma_b$, does not mediate a force between macroscopic objects.

This {\it induced gravity} approach requires a more elaborated discussion. It seems worthwhile to point out that, on the one hand, we exploit the induced gravity strategy not only in the E-frame, but also in the string one and, on the other hand, we are considering {\it multiple quantizations}. In other words, we quantized the theory more than once:\\

{\it Quantization - Step 1}. Our starting point in the S-frame is the strong-coupling stringy effective action \ref{3}, where the result of the non-perturbative quantum calculation is encoded in the form factors, which have been properly chosen to produce the lagrangian \ref{Ltotale}. Therefore, the Einstein-Hilbert term of formula \ref{eframe} is induced by a non-perturbative quantum calculation, namely, the first step.\\

{\it Quantization - Step 2}. We considered \ref{eframe} as a starting point for another quantization (see figure 1) and we discovered the presence of a conformal anomaly.\\

{\it Quantization - Step 3}. We exploited the anomaly induced interaction vertex \ref{vertice} to quantize once again the theory (see figure 2).\\

Since the dilaton controls the strength of all interactions, the
chameleonic behaviour of the field in the E-frame guarantees the
simultaneous presence of two different elements after step 3,
namely: (1) a (almost) negligible kinetic term for global
gravitons (therefore, global gravity is non-dynamical) and (2) a
non-negligible kinetic term for local gravitons obtained by
applying the induced gravity strategy for the first time (step 1).
The question is: {\it what kind of gravitational theory should we
expect locally?} At this stage, if we proceed with the
quantization {\it beyond} step 1, namely, if we apply the induced
gravity idea to the lagrangian \ref{eframe}, it is not yet clear
whether higher derivative gravitational terms are present locally.
However, this discussion is not directly relevant for the
cosmological constant problem and it will be addressed in a future
work. Moreover, in our proposal {\it all} the interactions are
switched off on cosmological length scales and it remains to be
seen whether this fact clashes with large-scale phenomenology. A
detailed phenomenological analysis is definitely necessary to
clarify these points.

\subsubsection{The chameleonic effective actions}
\label{azionieffettive}

In our proposal, the chameleonic behaviour
of the dilaton is compatible, {\it and it is the
effective result}, of a renormalization process which is carried
on to {\it all} orders in perturbation theory. The success of this
renormalization program is guaranteed by the restoration of
scale-invariance.

It must be stressed that in our approach we do {\it not} evaluate
the diagrams: we exploit the restoration of scale invariance on
cosmological distances to protect the mass of
the dilaton and the cosmological constant. Our
argument is valid {\it at all orders} in perturbation theory.

Remarkably, the chameleonic behaviour of the dilaton and the
splitting of the particles in global and local objects led us to a
formulation of the theory which strongly depends on the choice of
the length scale (parametrized by the value of the dilaton). This
comment must be further elaborated. Let us start by describing the
concept of {\it effective theory} in more detail. Let us consider
a particle physics theory, that we will call {\it fundamental},
where the degrees of freedom (particles and fields) are described
by a Lagrangian $L_A$. Moreover, let us consider a certain energy
scale $M$ and let us suppose that $L_A$ is able to correctly
describe physical processes not only below $M$, but also above
$M$. Now, if our intention is to describe physical phenomena {\it
below} $M$, we are free to describe these phenomena not only with
the lagrangian $L_A$, but also with another lagrangian $L_B$, that
we will call {\it effective} and that is obtained from the
fundamental one by integrating out the heavy degress of freedom.
In other words, the particles with a mass $m>M$ will be removed
from the effective theory and $L_B$ will be able to correctly
describe physical phenomena below (but not above) the energy scale
$M$. If we are interested in the description of the physical
phenomena above the scale $M$, then the reintroduction of the
heavy degrees of freedom will be mandatory. A well-known example
is given by the SM (theory A) and the Fermi theory (theory B).

Let us now come back to the chameleon. As already mentioned above,
we parametrize the length scales of the model through the dilaton
$\sigma$. In particular, as we argued above, large values of
$\sigma$ correspond to large distance scales (i.e. small mass
scales). In this way we can construct two different lagrangians
$L_A$ and $L_B$. On the one hand, the {\it local} particles can be
considered as the degrees of freedom of a lagrangian $L_A$,
describing physical phenomena below an energy scale $M_A$ fixed by
$\sigma_f$. On the other hand, {\it global} particles can be
considered as the degrees of freedom of a lagrangian $L_B$,
describing physical phenomena below an energy scale fixed by
$\sigma_b$ that we call $M_B \simeq H_0 <M_A$. $L_B$ can be
considered as the {\it effective theory} valid in the deep
infrared (IR) region, obtained from the (more fundamental)
lagrangian $L_A$ by integrating out the heavy degrees of
freedom\footnote{It seems now worthwhile to point out that in our
model the matter particles are {\it not} chameleons and
they can be safely integrated out on cosmological distances.}.  We
infer that the running of the dilaton towards large values can be
interpreted as the running of the theory towards the IR region
and, in general, a shift of the dilaton towards larger (smaller)
values will correspond to integrating out (in) degrees of freedom
through the exponential prefactor $e^{-\zeta\sigma}$ already
mentioned above. It is time to summarize our chameleonic
lagrangians (in the E-frame).

The local one is written exploiting {\it only local particles} as:
\begin{equation}
 L_{A}=\sqrt{-g_f^*}\left( \frac{1}{2} (R_*)_f -
    \half (g^{\mu\nu}_*)_f\partial_{\mu}\sigma_f\partial_{\nu}\sigma_f + {\cal L}_{* matter} + V_{eff}(\sigma_f)+ ...
    \right)
\label{eframenuova}
\end{equation}
Here the dots include the gauge part of the theory and they might
also include higher derivative (local) gravitational terms.
$V_{eff}(\sigma)$ is the total chameleonic effective potential for
$\sigma$ obtained by adding together (1) a run-away exponential
branch and (2) a matter branch linear in $\sigma$, given by
formula \ref{vertice}. ${\cal L}_{* matter}$ is given by:
\begin{equation}
{\cal L}_{* matter}= -
    \half (g^{\mu\nu}_*)_f D_{\mu}\Phi^*_f D_{\nu} \Phi_f^* - \xi^{-1} \frac{f}{4} M_p^2 (\Phi_f^*)^2+ \frac{\lambda_{\Phi}}{4!} (\Phi_f^*)^4
\label{lmatternuova}
\end{equation}
and $D_{\mu}=\partial_{\mu}+ \zeta \partial_{\mu} \sigma_f$. Needless to say, $L_A$ is not scale invariant.

The global lagrangian is written exploiting only the background
part of the fields and it is (almost) scale invariant. It is
written as:

\begin{equation}
 L_{B}=\sqrt{-g_{FRW}^*}\left( -\half (g^{\mu\nu}_*)_{FRW}\partial_{\mu}\sigma_b\partial_{\nu}\sigma_b + V_{eff}(\sigma_b)+ ...
    \right),
\label{eframenuova2}
\end{equation}
where the dots represent kinetic terms for massless global gauge fields. Therefore, the effective lagrangian of the model on cosmological distances corresponds (basically) to a free cosmological dilaton $\sigma_b$ in a FRW-background.  A promising line of development will analyze further the consequences of this IR-free theory in Cosmology and gravitational Physics (also from the standpoint of the AdS/CFT correspondence).

\subsection{The correct Dark Energy scale}
\label{thecorrect}

Naturally, our intention is to recover the correct Dark Energy
scale. We will exploit the chameleonic behaviour of the dilaton.
In the minimum of the effective potential we can write \beq
V(<\sigma_b>)= \rho_m B(<\sigma_b>), \eeq where $\rho_m$ is the
matter energy density and $B(\sigma)$ is the usual function of a
chameleonic model where the coupling is encoded. As previously
mentioned in the Introduction, Dark Matter and Dark Energy give a
similar contribution to the cosmic energy budget today.
Consequently, the correct DE scale is given by the chameleon
mechanism, granted that $B\simeq 1$. In the case of the coupling
\ref{vertice}, we identify the matter energy density in the
E-frame with $\frac{1}{2}m^2 \tilde{\Phi}^2$. Therefore we have
\beq B= \frac{1}{M_p \pi^2} \zeta \lambda \sigma \eeq and if we
choose $\lambda \simeq \zeta \simeq 1$, the condition $B=1$
requires $\sigma \simeq 1$. In the absence of a detailed analysis,
we have no theoretical grounds to support this value of $\sigma$.
Happily, however, this choice of parameters and fields is not a
fine-tuning. Remarkably, our B-function depends on $\sigma$ in a
{\it linear} way and, therefore, we can tolerate a certain
deviation of the value of $\sigma$ in the minimum from the
planckian scale.

\section{Discussion: the cosmological constant and non-equivalent frames}
\label{forum}

The careful reader may be puzzled by some of our considerations
and may ask a number of legitimate questions.\\

{\it What about a vanishing backreaction for the dilaton?}
This would clash with the non-linear nature of a chameleonic
theory.\\

{\it Why does the backreaction cancel the
unrenormalized vacuum energy?} A different result would clash with
the restoration of scale-invariance at large distances.
Accordingly, the effective potential must fall to {\it zero} on
cosmological distances and this result is obtained taking into
account all the quantum corrections. Backreaction is a crucial
element to keep under control the contribution of the quantum
diagrams to the cosmological constant, but this is true only for
the diagrams containing the dilaton and the gravitons.\\

{\it What
about a small value of $\zeta \sigma_b$? This would render the
suppression negligible.} The exponential suppression is a crucial
ingredient to restore scale invariance cosmologically and it
depends on the value of $\zeta \sigma$. In the chameleonic
literature an upper bound on the product $\zeta \sigma_{today}$ has
been discussed in connection with Big Bang Nucleosynthesis (BBN)
constraints on the variation of particle masses, see for example
\cite{Brax:2004qh}. However, these bounds cannot be directly
applied to our proposal, because our coupling function $B(\sigma)$ is linear. Therefore, it
is not the usual exponential function exploited to discuss the bounds. This difference could play an
important role in comparing the model with experimental
constraints. At this stage, the possibility still exists that a
certain amount of fine-tuning may be required to satisfy BBN
constraints. A detailed phenomenological analysis is definitely
necessary to clarify this point.\\

{\it This approach to the
cosmological constant problem is effective granted that you assume
the presence of scale-invariance in the Lagrangian.} As far as the S-frame larangian is concerned,
all that is really needed is a proper structure of the
S-frame form factors in the strong coupling regime. This will
guarantee the presence of a certain {\it sector} in the S-frame
Lagrangian where the symmetry is present. We can start with a
S-frame Lagrangian where scale-invariance is explicitly broken and
after the conformal transformation to the E-frame scale-invariance
will be restored (or, strictly speaking, established) on
cosmological distances. Consequently, the {\it global} Einstein frame lagrangian is (almost) scale-invariant.
In our scenario, the string dilaton in the
E-frame is the order parameter of the breaking of
scale-invariance. Since the dilaton is a chameleon, Particle
Physics will be the standard one only {\it locally}: the usual
contributions to the vacuum energy (e.g. from SUSY breaking, from
electroweak symmetry breaking, from axions...) will be extremely
large {\it locally}, but exponentially reduced on large
distances.\\

{\it Are the Jordan and Einstein frames
equivalent?} In this model the dilaton and the $\Phi$-field have
constant values in the S-frame and this is due to the presence of
a stringy lagrangian which is the result of a {\it quantum}
calculation. Needless to say, the cosmological constant in the
S-frame is much larger than the meV-scale. However, after the
conformal transformation to the E-frame the cosmological constant
is surprisingly under control (at the quantum level). These
comments are pointing out a {\it non}-equivalence of different
conformal frames at the quantum level and, in particular, the
E-frame is selected to be the physical frame. We warn the reader that we did not evaluate the cocycle function
(for an introduction see, for example, \cite{Kirsten:2001wz}),
because the cocycle is basically a conformal anomaly, therefore,
it is globally harmless and it will not clash with the restoration
of scale invariance on cosmological distances in the E-frame. We
infer that this correction does not modify the qualitative
characteristics of the chameleonic potential in the E-frame and it
can be neglected\footnote{In principle, there is also the
possibility that the cocycle will restore scale invariance
locally. In this case, whatever will be the length scale we choose
in the problem, the scale symmetry will be present. However, in
the absence of theoretical motivations to support this
possibility, we think that a locally non-symmetric configuration
is more generic.}.\\

{\it Is it possible to avoid the conformal transformation?
Can we make our calculations in one single (and physical) frame
(namely the Einstein one)?} The string frame is the natural frame
in which the string effective action is written and we need a conformal
transformation to establish a connection between string theory and
the physical (Einstein) frame. Therefore, it is not surprising or
strange to make a conformal transformation as an intermediate step
of our procedure: string theory is suggesting us to use a
non-physical frame (i.e. the string one).\\

{\it Do we have a dynamical mechanism that drives the
($\phi,\Phi$) system towards the equilibrium position in the
S-frame?} No. If the initial condition of the system is
characterized by non-fine-tuned values $\phi_0$ and $\Phi_0$, then
it is easy to choose the parameters of the model to satisfy the
field equations with those particular values of the
fields.\\

{\it Is this a successful chameleon model?}
A detailed phenomenological analysis is necessary to clarify
this point and it will be discussed in a future work. However,
in this paper we point out that, happily, in our model, the
chameleonic force cannot modify the motion of planets. This
point we mentioned last needs to be further elaborated.  As already
mentioned in the Introduction, a direct coupling between matter
and an ultralight scalar field can be phenomenologically
dangerous. Indeed, local limits on the coupling between matter and
a scalar field are particularly stringent and they require a small
coupling unless the scalar field is sufficiently heavy (i.e.
stabilization). Typically, one requires the (local) mass of the
chameleon to be larger than $10^{-3}$ eV. This condition can be
satisfied in a high density environment such as the atmosphere,
but in the solar system, where the density is much smaller, the
chameleon can be very light, it can mediate a long-range force and
it may be responsible for an (unacceptable) distortion of
planetary trajectories. To overcome this problem, the standard approach
discussed in the chameleonic literature is to exploit the so
called {\it thin-shell} mechanism. A body is said to have a
thin-shell if a chameleon $\chi$ is approximately constant
everywhere inside the body apart from a small region near the
surface of the body. Large ($\mathcal{O}(1)$) changes in the value
of $\chi$ can and do occur in this surface layer or thin-shell.
Inside a body with a thin-shell $\vec{\nabla}\chi$ vanishes
everywhere apart from a thin superficial layer. Since the
chameleonic force is proportional to $\vec{\nabla}\chi$, it is
only that surface layer, or \emph{thin-shell}, that both feels and
contributes to the `fifth force' mediated by $\chi$.
Therefore, when the thin-shell mechanism is operative,
the chameleon force between the Sun and the planets is very weak
and the otherwise tight limits on such a long-range force are
evaded \cite{Khoury:2003aq, Khoury:2003rn}. Remarkably, in a
recent paper \cite{Brax:2010gi}, the chameleonic behaviour of a
string dilaton running towards the strong coupling region has been
ruled out for an exponential coupling, because the thin-shell
mechanism was absent and unacceptable deviations from standard
gravity were predicted by the model. Let us now come back
to our model and let us discuss the motion of planets.
The anomaly induced interaction vertex \ref{vertice}
between dilaton and matter has been exploited to construct the
mass correction diagram of figure 2b, which is responsible for
the stabilization of the field. The exponential factor $e^{-\zeta \sigma}$ plays
a crucial role, on the one hand, to suppress the coupling to matter, on the other hand,
to suppress the mass of the dilaton. Since in our model there is no cosmological constant fine-tuning, the mass scale in the potential is planckian and the exponential suppression must produce a hierarchy of many orders of magnitude to obtain a long-range force. We infer that in our model the chameleonic dilaton $\sigma$ is ultralight {\it if and only if} it is non-interacting with matter. Consequently, the dangerous set up given by a very light scalar field with a relevant coupling with matter is naturally avoided in our proposal and, for this reason, the planetary orbits are not modified by our chameleon.

\section{Conclusions}

In this paper the chameleonic behaviour of the string dilaton has
been suggested and some of its consequences have been discussed in
detail. In particular (1) we proposed a new stringy solution to
the cosmological constant problem and (2) we pointed out a
non-equivalence of different conformal frames at the quantum
level. The correct Dark Energy scale is recovered in the E-frame
without fine-tunings of the parameters and this result is robust
against all quantum corrections, granted that we assume a proper
structure of the S-frame form factors in the strong coupling
regime. However, at this stage, the possibility still exists that
a certain amount of fine-tuning may be required to satisfy some
phenomenological constraints. Moreover, the theory is IR-free,
while higher derivative gravitational terms might be present locally
and it remains to be seen whether these facts clash with phenomenology.
A detailed phenomenological analysis
is definitely necessary to clarify these
points.

In our approach to the cosmological constant problem, the string
dilaton in the E-frame is the order parameter of the breaking of
scale-invariance. Since the dilaton is a chameleon, Particle
Physics will be the standard one only {\it locally}: the usual
contributions to the vacuum energy (e.g. from SUSY breaking, from
electroweak symmetry breaking, from axions...) will be extremely
large {\it locally}, but exponentially suppressed on large
distances.

To the best of our knowledge this was the first attempt of dealing
with the cosmological constant problem from the standpoint of
chameleon fields.

A chameleonic string dilaton, on the one hand, is a new stringy
way of tackling crucial problems like dilaton stabilization and
cosmological constant problem, on the other hand, it is a major
step forward to establish a connection between String Theory and
current experiments (most notably GammeV \cite{Chou:2008gr,
Upadhye:2009iv}).

We conclude by summarizing some of the possible lines of
development. One interesting project will study this model in the
framework of AdS/CFT correspondence and its possible connections
with the braneworld model of reference \cite{Zanzi:2006xr}.
Moreover,
the theory that we discussed in this paper and its
phenomenological consequences should be further investigated, on
the one hand, in Cosmology and gravitational physics (gravity is
globally absent and we must be sure that this fact does not clash
with phenomenology), on the other hand, in connection with the BBN
constraints on $\zeta \sigma$ (checking whether they can be faced
without fine tuning). The dual nature of the concept of particle
should be carefully investigated starting from
\cite{Colosi:2004vw, Buchbinder:1992rb, DeWitt:2003pm} and
references therein. The potential presence of higher derivative
gravitational terms in the local lagrangian should be also studied
considering the role played by multiple quantizations. One more
future research direction should be mentioned. In our model a
stabilized dilaton in the S-frame does not correspond to a
stabilized dilaton in the E-frame and, at first glance, this
result seems to clash with the approach discussed in
\cite{Zanzi:2006xr}. These issues will be discussed in a future
work.

\subsection*{Acknowledgements}

Special thanks are due to Antonio Masiero and Massimo Pietroni for
valuable discussions. I warmly thank Yasunori Fujii and Maurizio Gasperini for useful
correspondence. I acknowledge support from "Angelo della Riccia"
foundation.


\providecommand{\href}[2]{#2}\begingroup\raggedright\endgroup

\end{document}